\documentclass[aps,prl,reprint,twocolumn,showpacs,superscriptaddress]{revtex4-2}
\usepackage{epsfig}
\usepackage{natbib}
\usepackage{ulem}
\usepackage{xspace}
\usepackage{xcolor}
\usepackage{graphicx}
\usepackage{amssymb}
\usepackage{cancel}
\usepackage{ulem}
\usepackage{lineno}
\usepackage{amsmath}
\usepackage{stackrel}
\usepackage{appendix}
\usepackage{url}
 \usepackage{enumitem}
 \usepackage{hyperref}

\hypersetup{
	colorlinks=true,
	linkcolor=blue,
	citecolor=blue,
	urlcolor=blue
}

\newcommand{\odd}{\mathrm{odd}}
\newcommand{\even}{\mathrm{even}}

\newcommand{\I}{\mathcal{I}}
\newcommand{\Iodd}{\mathcal{I}_\mathrm{odd}}

\begin{document}    

\title{Stretched-exponential relaxation in weakly-confined Brownian systems through large deviation theory}

		\author{Lucianno Defaveri}
        \affiliation{Department of Physics, Bar-Ilan University, Ramat-Gan 52900, Israel}
		\author{Eli Barkai}
        \affiliation{Department of Physics, Institute of Nanotechnology and Advanced Materials, Bar-Ilan University, Ramat-Gan 52900, Israel}
		\author{David A. Kessler}
        \affiliation{Department of Physics, Bar-Ilan University, Ramat-Gan 52900, Israel}

\begin{abstract} 
  	Stretched-exponential relaxation is a widely observed phenomenon found in ordered ferromagnets as well as glassy systems.
    One modelling approach connects this behavior to a droplet dynamics described by an effective Langevin equation for the droplet radius with a $r^{2/3}$ potential.
    Here, we study a Brownian particle under the influence of a general confining, albeit weak, potential field that grows with distance as a sub-linear power law.  
    We find that for this memoryless model, observables display stretched-exponential relaxation.
    The probability density function of the system is studied using a rate function ansatz. We obtain analytically the stretched-exponential exponent along with an anomalous power-law scaling of length with time. The rate function exhibits a point of nonanalyticity, indicating a dynamical phase transition. In particular, the rate function is double-valued both to the left and right of this point, leading to four different rate functions, depending on the choice of initial conditions and symmetry. 
\end{abstract}

\maketitle

\textit{Introduction.}-- Anomalous relaxation, characterized by non-exponential decay, is observed in a wide range of physical systems \cite{Uchaikin2013}. One class of such behavior is stretched-exponential relaxation \cite{Kohlrausch1854,Klafter1986,Kisslinger1993,Klafter2005,Mukherjee2023}. This has been seen, for example, in disordered or heterogeneous systems, where the complex interplay of interactions leads to a broad distribution of relaxation times~\cite{Malacarne2001}.
Moreover, the stretched-exponential behavior is observed in particle movement in disordered dielectrics \cite{Milovanov2007,Klafter2002,Weron2006} and in relaxation of glassy-dynamics~\cite{Kakalios1987,Wu2018}. 

Some studies of the relaxation of ordered Ising ferromagnets have sought, building on the work of Huse and Fisher~\cite{Fisher1987,Fisher1988}, to connect the
observed stretched-exponential relaxation to an effective droplet dynamics, which was mapped to
a Langevin equation with a $x^{2/3}$ potential~\cite{Langer1989}. It was shown that this model does indeed exhibit stretched-exponential relaxation.
In this Letter, we present a solution to the Fokker-Planck equation with a general external potential that grows with distance as a sub-linear power law. We show that for this class of models the relaxation of the various moments of the position to their equilibrium values follows an anomalous stretched exponential. 
Formally, the Fokker-Planck equation (see Eq.\,(\ref{eq:fokker-planck})) for $P(x,t)$, the probability density function (PDF), can be solved via an eigenfunction expansion, and yields stretched-exponential relaxation.
However, it turns out that there is a simpler, more direct, approach, which also yields additional physical insight.  We find that in the long-time limit, $P(x,t)$ takes the form of the exponential of a rate function which possesses a scaling form, as usually results from a large-deviation formalism~\cite{Freidlin1998}, which looks at the far tails of the distribution of an observable. Large deviation theory is a subject of much active interest in statistical physics,~\cite{Touchette2009,Kafri2015,Kafri2018,Touchette2018,Meerson2019,Touchette2016,Touchette2017,Harris2020,Espigares2023a,Oren2023}, in particular, due to the recent discovery of dynamical phase transitions in the large-deviation behavior of some model systems~\cite{Janas2016,Nemoto2017,Smith2022,Smith2022a,Smith2022b,Smith2023,Espigares2023b,Agranov2023}. 
These are nonanalytic points of the rate function, and are so-called due to the analogy of the rate function to an equilibrium free energy \cite{Touchette2016,Touchette2018, Smith2022}.
We show the relationship between the stretched-exponential relaxation and the presence of a dynamical phase transition.

The rate function is defined as the logarithm of the PDF $P(x,t)$, divided by a power of the time~\cite{Touchette2018,Meerson2019,Smith2022}
\begin{eqnarray}
    \I(z) \equiv  \hspace{-0.15in} \lim_{\stackrel{t,x\to\infty}{z=x/t^\gamma \textrm{fixed}}} \hspace{-0.1in}\frac{-\ln P(x,t)}{t^\nu} \, ,
    \label{eq:Iscaling}
    \label{eq:rate-function-def}
\end{eqnarray}
with the anomalous time-exponent $\nu \neq 1$ and where $\I$ is a function of the scaling variable $z\equiv x/t^{\gamma}$. The stretched-exponential relaxation of observables will be governed by the same anomalous exponent $\nu$.  It should be noted that in the previously identified cases with anomalous temporal scaling, the observable in question was nonlocal in time, whereas here it is the PDF of $x$ itself that exhibits the anomalous scaling. The appearance of a rate function in our problem implies the surprising result that the anomalous stretched-exponential relaxation we observe is a result of large deviations, i.e. the dynamics at large $x$.

We find that the rate function is multivalued, and possesses a critical-point $z_c$.
The rate function has two possible branches for $z<z_c$ and two for $z>z_c$, all meeting at $z_c$. All four possible combinations of branches below and above $z_c$ have different interpretations, corresponding to different classes of initial conditions and parity. Two of the combinations have a jump discontinuity in $\I''(z)$, indicating the presence of a dynamical phase transition (see \cite{Touchette2016, Smith2022} for more details). Such a multi-valued rate function appears not to have previously been encountered. Using the appropriate branches of the rate function, we obtain the time scales of the stretched-exponential relaxation of the even and odd moments of $x$.

\textit{Model.}-- We study non-interacting Brownian particles, in contact with a heat bath at temperature $T$, that are also subject to a binding potential field $V(x)$.  The spatial spreading of the cloud of particles can be described via the PDF $P(x,t)$, obtained from the Fokker-Planck equation  (FPE)
\begin{eqnarray}
\frac{\partial}{\partial t} P(x,t) = D \left[ \frac{\partial^2}{\partial x^2} + \frac{\partial}{\partial x} \left( \frac{V'(x)}{k_B T}  \right) \right] P(x,t) \, , \label{eq:fokker-planck}
\end{eqnarray}
where $D$ is the diffusion coefficient. We consider herein even potentials, satisfying $V(x) = V(-x)$, which for large $x$ grow as a sublinear power-law, $V(x) \propto x^\alpha$, where $0 < \alpha < 1$. This means that the force, $F(x) = - V'(x)$,  will be negligible for large $x$, as $F(x) \propto x^{\alpha-1}$. At long times, the particles will reach the stationary equilibrium Boltzmann-Gibbs state $P_\mathrm{BG}(x) = \exp [ - V(x)/k_B T]/Z$, with $Z$ being the normalizing partition function. For the more commonly considered superlinear growth of the potential, $\alpha>1$, $F(x)$ grows with distance and everything is standard. The system exponentially relaxes to the Boltzmann-Gibbs state, at a rate given by the first nonzero eigenvalue of the Fokker-Planck operator, which has a discrete spectrum starting at 0. This discreteness follows from the fact that under a similarity transformation, the FPE becomes a Schr\"odinger equation \cite{Risken1996} with an effective potential $V_{S}(x)=F(x)^2/4k_B T + F'(x)/2 k_B T$, which grows without bound as $x\to \infty$. However, for $\alpha<1$, $V_{S}(x)$ decays at large $x$ as $x^{-2(1-\alpha)}$ and so the spectrum goes continuously down to 0. 
Potentials of this form have already been studied in the context of resetting processes \cite{Capala2023} and active processes \cite{Dhar2019}, in particular, the $\alpha=1$ limiting case \cite{Chen2013,Majumdar2020,Zarfaty2022}.
As noted above, in Ising ferromagnets below the critical temperature, the dynamics of spherical droplets have been modelled using an effective potential which, for $d=3$ dimensions, is equivalent to $\alpha=1$ and, for $d=2$ dimensions, is equivalent to $\alpha=2/3$~\cite{Fisher1987,Fisher1988,Langer1989}. For $\alpha \to 0$ (and assuming $V(x) \sim x^\alpha / \alpha$), we have that for large $x$, $V(x) \sim \ln x$. In that limit, the relaxation has been shown to be governed by a power law \cite{Kessler2010,Dechant2011,Mukamel2011,Mukamel2012}. The question is then what happens for $0 < \alpha < 1$?  

For our numerical examples, we use the family of potentials
\begin{eqnarray}
V(x) = V_0 \left( 1 + \frac{x^2}{\ell^2} \right)^{\alpha/2} \, . \label{eq:def-pot}
\end{eqnarray}
For $x \gg \ell$, with $\ell$ being the lengthscale of the center region, the potential exhibits the desired power law growth. Throughout this Letter, we shall scale the position variable by $\ell$  and correspondingly the time by $\ell^2/D$.

As mentioned above, the $P(x,t)$ can be decomposed in a sum of eigenfunctions, each decaying as $e^{-\lambda t}$ with a continuous eigenvalue spectrum $\lambda$ starting at $0^+$, together with the Boltmann-Gibbs bound state at $\lambda=0$.
It turns out that the dominant continuum contribution to $P(x,t)$ for large times comes from the vicinity of a particular finite eigenvalue $\lambda_*$.
This dominant eigenvalue scales as a negative power of the time, $\lambda_*(t)\sim t^{\nu-1}$, $0\le\nu<1$, so that $e^{-\lambda_*(t) t}~\sim e^{-t^\nu}$, a stretched-exponential relaxation to equilibrium. This stretched-exponential relaxation is easily demonstrated numerically (see below). The eigenvalue calculation will be sketched in the Supplemental Material (SM) \cite{sup_mat}. We turn now instead to the rate-function calculation.  Most strikingly, we shall see that the power of $t$ in the stretched exponential can be obtained immediately from our scaling ansatz for the rate function.

\begin{figure}
    \centering
    \includegraphics[width=0.45\textwidth]{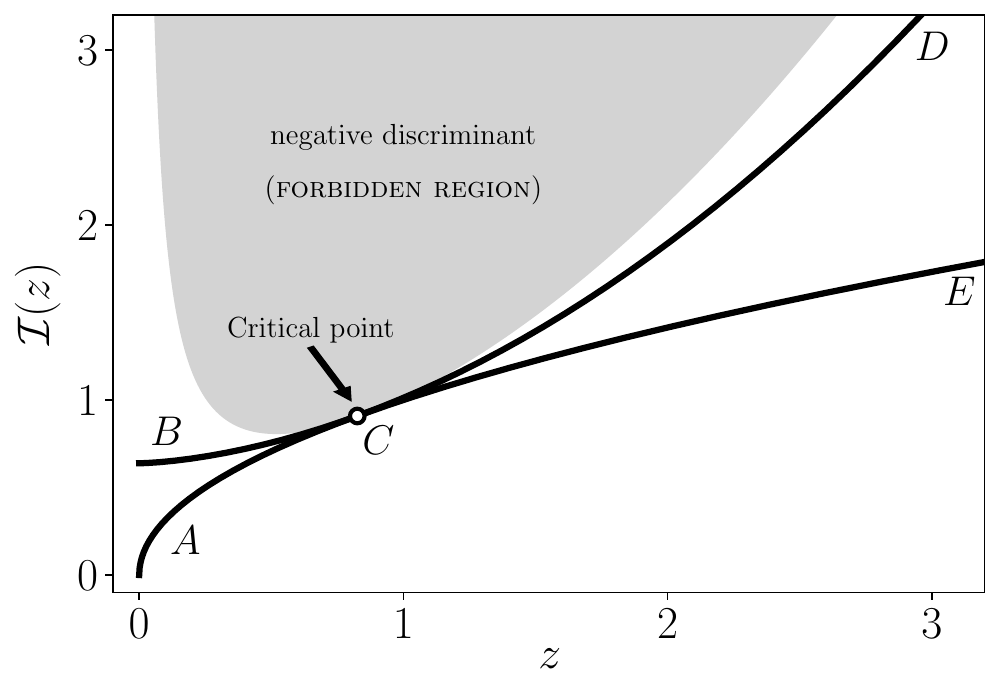}
    \caption{All possible rate functions, defined by Eq.\,(\ref{eq:rate-function-def}), which are solutions of the differential equation in Eq.\,(\ref{eq:df-rate-function}), versus the scaled position $z \equiv x/t^\gamma$ (the scaled exponents defined in Eq.\,(\ref{eq:scaled-exponents})) for $\alpha=1/2$. The rate function has two low-$z$ and two high-$z$ branches, leading to four different possible forms. The region where the discriminant of Eq.\,(\ref{eq:branches-Iprime}) is negative is highlighted in gray.}
    \label{fig:rate-functions}
\end{figure}

\textit{Large deviation formalism}.-- 
From Eq.\,(\ref{eq:rate-function-def}) we can write the PDF, up to pre-exponential factors, as $P(x,t) \sim e^{-t^\nu \I(z)}$. Inserting this ansatz into the FPE (\ref{eq:fokker-planck}), we find that in the long-time limit
\begin{eqnarray}
\frac{z \gamma \I'(z) - \nu \I(z)}{t^{1-\nu}} = \frac{\I'(z)^2}{t^{2 \gamma - 2 \nu}} - \frac{V_0}{k_B T}\frac{ \alpha }{z^{1-\alpha}} \frac{\I'(z)}{t^{(2-\alpha)\gamma - \nu}} .\label{eq:long-time-scaling}
\end{eqnarray}
We impose that the time dependence of all terms is the same, namely, $1-\nu = 2 \gamma - 2 \nu = (2-\alpha)\gamma - \nu$, and find that the scaling exponents are
\begin{eqnarray}
\nu = \frac{\alpha}{2-\alpha} ~ \mathrm{and} ~ \gamma = \frac{1}{2-\alpha} \, . \label{eq:scaled-exponents}
\end{eqnarray}
With these exponents, both the exponent of the Boltzmann-Gibbs solution, $V(x)/k_B T = V_0/k_B T x^\alpha  = t^\nu V_0/k_B T z^\alpha$, and of free-diffusion, $x^2/4Dt = t^\nu z^2/4D$, are compatible with our scaling. 

The rate function can be found by solving the non-linear differential equation
\eqref{eq:long-time-scaling}, which now reads
\begin{eqnarray}
\mathcal{I}'(z)^2 - \left( \frac{V_0}{k_B T} \frac{\alpha}{
z^{1-\alpha}} + \frac{z}{2-\alpha} \right) \mathcal{I}'(z) + \frac{\alpha \mathcal{I}(z)}{2-\alpha} = 0 . \label{eq:df-rate-function}
\end{eqnarray}
When we consider the large-$z$ limit, the rate function solution will assume the form $\I(z) \approx \xi_0 z^{\mu_0}$, where $\mu_0$ and $\xi_0$ must be determined. Plugging $\I(z)$ into Eq.\,(\ref{eq:df-rate-function}) in the limit of large $z$, we have
\begin{eqnarray}
    \xi_0 \mu_0^2 z^{\mu_0-2} - \frac{V_0}{k_B T} \alpha \mu_0 z^{\alpha - 2}  \sim \left( \frac{\mu_0 - \alpha}{2 - \alpha} \right) \, .
\end{eqnarray}
There are two possible solutions for this equation. First, we have $\mu_0=\alpha$, leading to $\I(z) = V_0 z^\alpha / k_B T$, i.e. the aforementioned Boltzmann-Gibbs state, which is a solution of Eq.\,(\ref{eq:df-rate-function}) for all $z$. Second, we have $\mu_0=2$ and $\xi_0 = 1/4$, which is equivalent to large $z$ diffusive behavior.  We emphasize that these are the only possible asymptotic solutions of Eq.\,(\ref{eq:df-rate-function}). Both these behaviors are necessary.  The Boltzmann-Gibbs state is a possible time-independent solution of the problem, when the initial state is the equilibrium state. However, under any initial conditions that decay with $x$ faster than the Boltzmann-Gibbs state and for any finite time $t$, we cannot expect the Boltzmann-Gibbs state to describe the whole PDF. The particles cannot spread faster than what is permitted by diffusion. Since $F(x) \to 0$ as $x\to\infty$, we expect diffusive behavior at large $|x|$, consistent with the second asymptotic behavior.

The next step is to solve Eq.\,(\ref{eq:df-rate-function}) globally.
Since Eq.\,(\ref{eq:df-rate-function}) is a second-order polynomial in ${\cal I}'(z)$, we have in fact two different options for the ODE:
\begin{align}
\I'(z) &= \frac{1}{2}\left( \frac{V_0}{k_B T} \frac{\alpha}{
z^{1-\alpha}} + \frac{z}{2-\alpha} \right)  \nonumber \\
& {}\pm \frac{1}{2} \sqrt{\left( \frac{V_0}{k_B T} \frac{ \alpha}{z^{1-\alpha}} + \frac{z}{2-\alpha} \right)^2 - \frac{4 \alpha \I(z)}{2 - \alpha}} \, . \label{eq:branches-Iprime}
\end{align}
These two ODEs give rise to two smooth solutions that cross at a critical point $z_c$, at which the square-root vanishes. One solution has the positive sign of the square root for $z<z_c$ and the negative sign for $z>z_c$ and the other with the opposite choices. Interestingly, two other solutions are also possible, where the sign does not switch, and which have a jump in ${\cal I}''$, leaving us with four possible descriptions, two for $z<z_c$ and two for $z>z_c$, as shown in Fig.\,\ref{fig:rate-functions}.  The next task is to uncover the physical content of these rate-function branches.

\textit{Boltzmann-Gibbs thermal initial condition}.--
As noted above, if we were to start at time $t=0$ with the thermal state in all of space, the state would remain unchanged for all time. In Fig.\,\ref{fig:rate-functions}, this is shown as the curve $ACE$.
The square-root in Eq.\,(\ref{eq:branches-Iprime}) vanishes at the critical point $z_c$,
\begin{eqnarray}
z_c = \left( \frac{\alpha(2-\alpha) V_0}{ k_B T} \right)^{1/(2-\alpha)} \label{eq:def-zc}  \, .
\end{eqnarray}
 The pure thermal state is obtained when we switch from positive ($z<z_c$) to negative ($z>z_c$) in Eq.\,(\ref{eq:branches-Iprime}).

\textit{Localized initial condition}.-- Our goal is to associate the solutions shown in Fig.\,\ref{fig:rate-functions} with classes of initial conditions and with parity. For an initially localized packet of particles, we expect the large $z$ behavior to be diffusive rather than Boltzmann-Gibbs. However, for small $z$, we expect the behavior to match Boltzmann-Gibbs, so that the Boltzmann-Gibbs regime in $x$ expands as $t^\gamma$. Keeping the positive sign of the square-root for $z>z_c$ leads to the desired diffusive-like large $z$ behavior.  The resulting singularity in ${\cal I}''$ is precisely the dynamical phase transition. The resulting curve, $ACD$ in Fig.\,\ref{fig:rate-functions}, describes the localized initial condition. 

\begin{figure}
    \centering
    \includegraphics[width=0.45\textwidth]{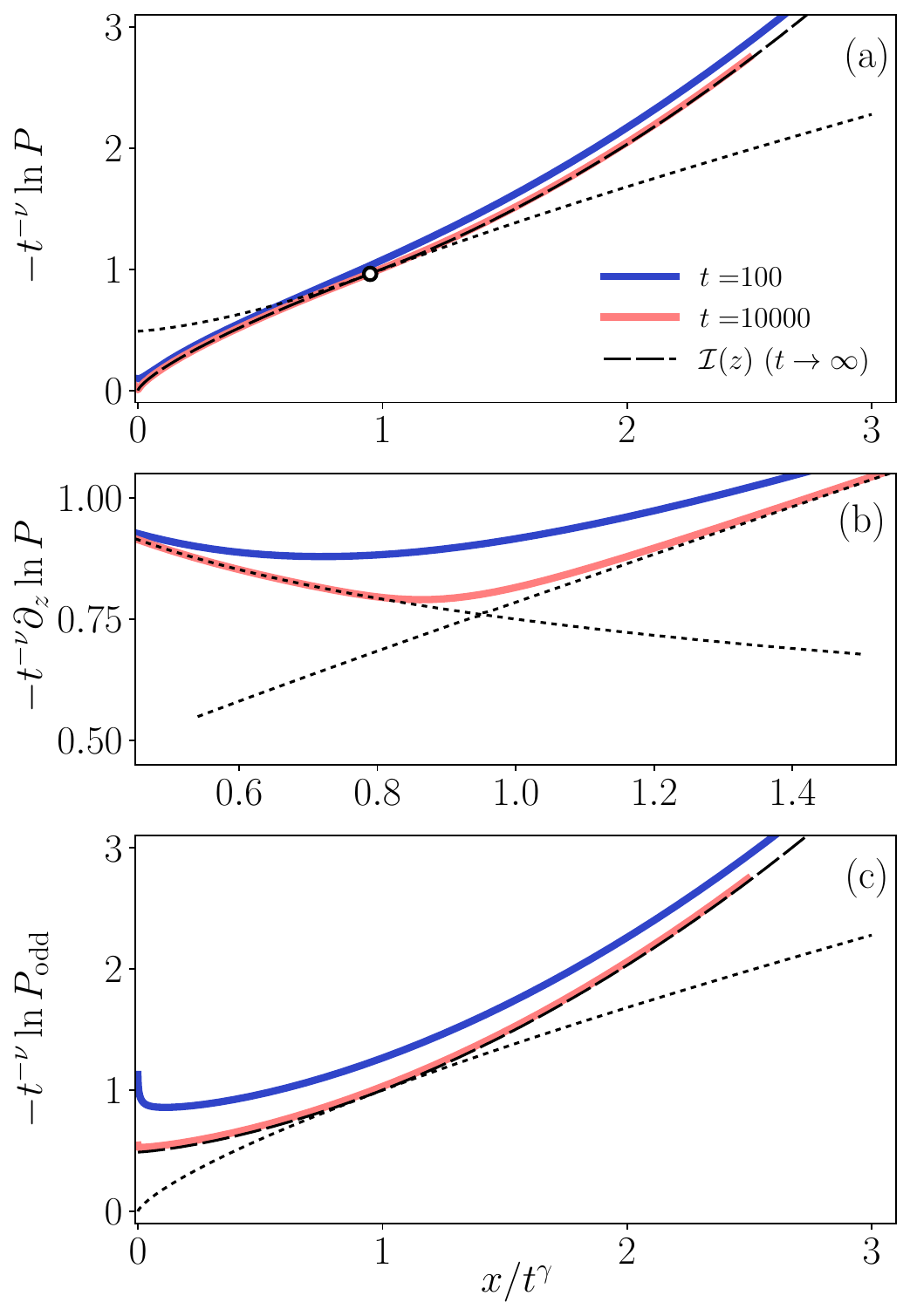}
    \caption{Comparison between theoretical and numerical (at finite times) rate functions, Eq.\,(\ref{eq:rate-function-def}),  as a function of the scaled position $z\equiv x/t^\gamma$. The solid lines represent numerical results obtained integrating Eq. (\ref{eq:fokker-planck}) at different times, indicated in the legend (panel (a)) for all panels. The dotted black lines represent all four possible rate functions (see  Fig.\,\ref{fig:rate-functions}) in panels (a) and (c) and their derivatives in panel (b).  In panel (a) we show the clear convergence to the rate function $\I_{ACD}(z)$ and highlight (white-filled circle) the critical transition point $z_c$. In panel (b) we show the derivative $\I'(z)$, and the non-analytical behavior becomes clear. The solution is equivalent to Boltzmann-Gibbs up until the critical point, where the system changes to the free-particle solution for larger values of $z$. In panel (c) we have the odd rate functions, where there is no dynamical phase transition. The numerical solutions demonstrate clear convergence towards the expected rate functions. We have used $\alpha = 3/4$, $\ell = 1$, $D=1$ and $V_0/k_B T = 1$.}
    \label{fig:rate-functions-compared-numerical}
\end{figure}

We can write odd/even PDFs using distinct rate functions as
\begin{eqnarray}
P_{\mathrm{even}} \sim e^{- t^\nu \mathcal{I}_\mathrm{even}(z)} ~ \mathrm{and} ~ P_{\mathrm{odd}} \sim e^{- t^\nu \mathcal{I}_\mathrm{odd}(z)} \, . \label{eq:I-odd-even}
\end{eqnarray}
If the particles start at the origin ($x_0=0$), then by symmetry $P_{\odd}(x,t) = 0$, since $V(x)$ is even and so all odd moments vanish, as they do in equilibrium. On the other hand, for $x_0 \neq 0$, the odd part is present, $P_{\odd}=(P(x,t)-P(-x,t))/2$, and is described by the curve $BCD$. We then write that $\I_\odd(z) \equiv \I_{BCD}(z)$, which tends to a constant value $\Iodd(0)$ as $z$ approaches zero, see Fig.\,\ref{fig:rate-functions}. Therefore, we have that $P_\odd(x,t)$ has an upper bound that decays as $e^{-\I_\odd(0) t^\nu}$, indicating that the odd contributions will decay as a stretched exponential. Curve $BCE$ describes the contribution from the continuum modes. This contribution is obtained by subtracting from $P(x,t)$ the Boltzmann-Gibbs solution, $P^* \equiv P - P_\mathrm{BG}$. At large $x$, this results in the negative of the Boltzmann-Gibbs solution, as the full solution is small here. The stretched-exponential relaxation of the even moments of $x$ to their equilibrium values is controlled by $P^*$. The rate function, $BCE$ in Fig.\,\ref{fig:rate-functions}, also displays a dynamical phase transition at the critical point. Finally, the even part, $P_{\even}=(P(x,t)+P(-x,t))/2$, is described by the $ACD$ rate function, $\I_\even(z) \equiv \I_{ACD}(z)$. This solution also captures the leading behavior of the density $P(x,t)$ for localized initial conditions.

We presented the four different scenarios for the rate function represented by branches in Fig.\,\ref{fig:rate-functions}. Which scenario is the relevant one is based on the choice of the initial condition and the parity. We now use finite time numerical integration of the FPE to show that, using Eq.\,(\ref{eq:rate-function-def}) (adapted for finite times), the results converge to the expected rate functions. In Fig.\,\ref{fig:rate-functions-compared-numerical}(a), for localized initial conditions, we show the numerical convergence to the rate function $ACD$, while in Fig.\,\ref{fig:rate-functions-compared-numerical}(b) we show the derivative of the curve $ACD$. The derivative clearly shows the non-analytical behavior at the point $z_c$. 
In Fig.\,\ref{fig:rate-functions-compared-numerical}(c), we compare our long-time prediction for $\Iodd(z)$ ($I_{BCD}(z)$) with numerical results ($t^{-\nu} \log P_\odd$, with $P_\odd$ obtained numerically), showing clear convergence.

In summary, we find that two of the rate functions are completely analytical. Those are the Boltzmann-Gibbs (curve $ACE$) and the odd rate functions (curve $BCD$). The other two possibilities, the localized initial condition rate function (curve $ACD$) and the continuum modes rate function (curve $BCE$) have a non-analytical behavior, and therefore a dynamical phase transition, at the critical point $z_c$. 

\textit{Remark on the notation and prefactors}.-- We highlight that the rate functions defined in Eq.\,(\ref{eq:I-odd-even}) satisfy $\I_{\even/\odd} (z) = \I_{\even/\odd} (-z)$ as these functions are even. Further, they do not depend explicitly on the initial conditions. The parity of $P_{\even/\odd}$ is determined by the pre-exponential factors. The exponential prefactors of $P_\mathrm{odd}$ will depend on the initial condition, since for symmetric initial conditions, the whole odd part must vanish. The prefactor is obtained using the WKB method in the SM.

\begin{figure}
    \centering
    \includegraphics[width=0.45\textwidth]{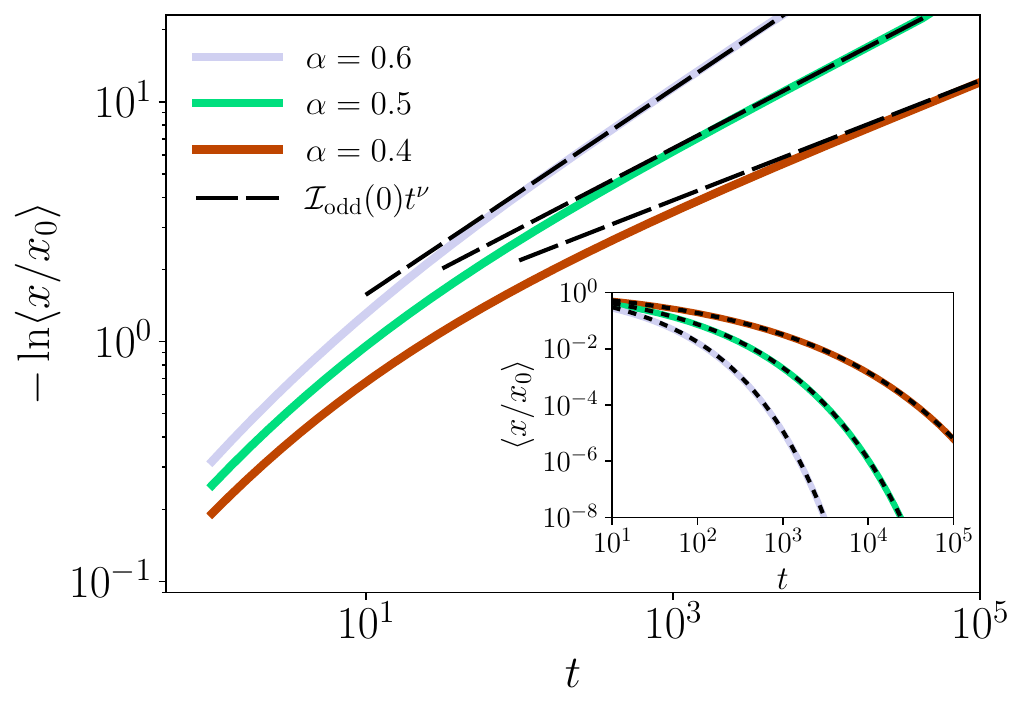}
    \caption{The log of the numerical ensemble average of the position for a system initialized with localized condition $P(x,t=0) = \delta(x-x_0)$ for different values of $\alpha$ (shown in legend). The stretched-exponential behavior is clearly shown for long times, where we compare with $\Iodd(0) t^\nu$ (dashed black lines), where the anomalous time scale $\Iodd(0)$ is obtained in Eq. (\ref{eq:anomalous-time-scale}), and $\nu = \alpha/(2 - \alpha)$.
    On the inset, the complete theoretical prediction (dotted black lines), described by Eq. (\ref{eq:mean-of-x-stretched}), is compared with the numerical simulations (colored lines) for the same exponents $\alpha$ as in the main label.  We have used $V_0/k_B T = 1$, $\ell = 1$ and $x_0=0.04$.}
    \label{fig:relaxation-mean}
\end{figure}

\textit{Anomalous relaxation}.-- Here we study the relaxation properties of the system, focusing on the first moment of the position $x$, starting with an asymmetric initial condition ($x_0 \neq 0$). Because this observable, $x$, is odd, the only non-zero contribution comes from the asymmetric part of the PDF. The odd part of the PDF can be written using $\Iodd$, up to pre-exponential factors, as shown in Eq.\,(\ref{eq:I-odd-even}).  We obtain, up to pre-exponential factors, the stretched-exponential characteristic of the mean as
\begin{eqnarray}
\langle x \rangle = \int_{-\infty}^\infty x \, P_\odd(x,t) dx \sim e^{-\Iodd(0) t^\nu} \, . \label{eq:mean-ensemble}
\end{eqnarray}
The value of $\Iodd(0)$ governs the anomalous time-scale $\tau$ of the relaxation. It is possible to obtain this value numerically by integrating Eq.\,(\ref{eq:branches-Iprime}) in the correct branch (positive sign). We have obtained an analytical expression through our eigenfunction calculations (see SM)
\begin{eqnarray}
\frac{1}{\tau^\nu} = \Iodd(0) = \frac{\left( \sqrt{\pi} \left( \frac{\alpha V_0}{2 k_B T} \right)^{\frac{1}{1-\alpha}} \frac{\Gamma\left( \frac{\alpha}{2 - 2 \alpha} \right)}{\Gamma\left( \frac{1}{2 - 2 \alpha} \right)} \right)^{1-\nu}}{\nu^\nu(1-\nu)^{1-\nu}} . \label{eq:anomalous-time-scale}
\end{eqnarray}
It is remarkable that a rate function controls the relaxation, in the sense that it cannot be considered describing a rare event, nor is it particularly hard to measure it. 
In Fig.\,\ref{fig:relaxation-mean}, the stretched-exponential behavior is shown numerically in the long-time limit. The observed time scale matches our prediction in Eq.\,(\ref{eq:anomalous-time-scale}) and is independent of the odd moment under consideration. The same can be extended for even moments $x^{2n}$ (see SM).
In order to obtain a complete expression for the mean in Eq.\,(\ref{eq:mean-ensemble}), the rate function is not enough. We must  account for the pre-exponential factor, $\mathcal{A}(t)$, that is, $P_\odd \approx \mathcal{A}(t) \, x^{1-\alpha} e^{- \Iodd(z)t^\nu}$.

In the long-time limit, the main contribution to the integral in Eq.\,(\ref{eq:mean-ensemble}) arises from the region where $x$ is much smaller than the critical $x_c(t)$, corresponding to the small $z$ region \cite{small_z_region_explanation}. For small $z$, we have $\Iodd(z) \approx \Iodd(0)(1  + k_B T z^{2-\alpha} / V_0 (2-\alpha)^2 )$, and the pre-exponential factor is (see SM)
\begin{eqnarray}
    \mathcal{A} (t) = x_0 \frac{k_B T}{\alpha V_0}  \frac{\Iodd(0) e^{\frac{V(0)}{k_B T}}}{t^{3/2-\nu}} \sqrt{\frac{\gamma - \nu}{4 \pi}} \label{eq:prefactor-A} \, .
\end{eqnarray}
Note that for the transition value $\alpha = 1$, $\gamma = \nu = 1$, and $\mathcal{A} (t)$ will be null. With the contribution of the prefactor, we obtain, from Eq.\,(\ref{eq:mean-ensemble}), the long-time expression for the relaxation,
\begin{eqnarray}
\frac{\langle x \rangle}{x_0} & \sim & \frac{\mathcal{C}_1 \left( \frac{V_0}{k_B T}, \alpha \right)}{t^{\nu^2/2}} e^{- \Iodd(0) t^\nu} \label{eq:mean-of-x-stretched} \, ,
\end{eqnarray}
where the definition of $\mathcal{C}_1 \left( {V_0}/{k_B T}, \alpha \right)$ is found in the SM.
Thus, the time relaxation of the system to equilibrium is through a stretched exponential (multiplied by a power law in time). We show the excellent agreement of Eq.\,(\ref{eq:mean-of-x-stretched})  with the numerical results in the inset of Fig.\,\ref{fig:relaxation-mean}. 

\textit{Discussion}.-- The results we obtained in this Letter are quite general, but nevertheless, many extensions to this work are possible. As a first step in this direction, we studied a case in dimensions higher than one, showing that the main results remain valid. The characteristics of time-averaged observables and the potential link between different branches and singularities in the cumulant generating function warrant attention \cite{Stella2023,Stella2023b}. It is likely that the multivalued nature of the rate function under study, which depends on symmetry and initial condition, is an important feature for other systems. Thus, the appearance of the multivalued rate function in more systems and its relationship with dynamical phase transitions requires further study.

{\bf Acknoledgements:}
The support of Israel Science Foundation's grant 1614/21 is acknowledged. LD and EB thank Naftali Smith for the insightful conversations.

%


\begin{thebibliography}{48}
\bibitem [{sma()}]{small_z_region_explanation}%
	\BibitemOpen
	\href@noop {} {}\bibinfo {note} {The maximum contribution to the integral in
		Eq.\,(\ref{eq:mean-ensemble}) is centered at $x^*(t)$. In the long-time
		limit, $x^*(t)/x_c(t) \equiv z^*(t)/z_c(t) \sim t^{-\gamma \nu} \to 0$, and
		therefore, the important region is the $z \to 0$ region.}\BibitemShut {Stop}%
\end{thebibliography}

\end{document}


\title{Suplemental Material for: Stretched-exponential relaxation in weakly-confined Brownian systems through large deviation theory}

		\author{Lucianno Defaveri}
        \affiliation{Department of Physics, Bar-Ilan University, Ramat Gan 52900, Israel}
		\author{Eli Barkai}
        \affiliation{Department of Physics, Institute of Nanotechnology and Advanced Materials, Bar-Ilan University, Ramat Gan 52900, Israel}
		\author{David A. Kessler}
        \affiliation{Department of Physics, Bar-Ilan University, Ramat Gan 52900, Israel}

\maketitle

\section{The eigenfunction expansion}

The Fokker-Planck equation can be solved formally through an eigenfunction expansion \cite{Risken}. The probability density function (PDF) at time $t$, $P(x,t)$, for localized initial conditions $P(x,t=0) = \delta(x-x_0)$ is
\begin{eqnarray}
    P(x,t) &=& e^{-V(x)/2 k_B T + V(0)/2 k_B T} \sum_k N_k^{-2} \psi_k (x_0) \psi_k(x) e^{- \lambda_k t} \, .
\end{eqnarray}
In the previous equation, $\lambda_k$ represents the $k$-th eigenvalue associated with the eigenfunction $\psi_k(x)$. Each eigenfunction is normalized by $N_k$. The eigenfunctions are obtained as the solutions of the Schr\"odinger equation \cite{Risken}
\begin{eqnarray}
    [V_S(x)  - \partial_x^2 ] \psi_k(x) = \lambda_k \psi_k(x) \, ,
\end{eqnarray}
where the effective potential is defined as
\begin{eqnarray}
    V_S(x) &=& \frac{V'(x)^2}{4 (k_B T)^2} - \frac{V''(x)}{2 k_B T} \, .
\end{eqnarray}
For the potentials we will consider herein, there always exists a thermal equilibrium state. Since this solution is time-independent, it is convenient to separate the ground state ($\lambda_0 = 0$) from the excited states as
\begin{eqnarray}
    P(x,t) = P_\eq(x) + P^*(x,t) \, . \label{eq:BG-excited}
\end{eqnarray}
The equilibrium state is
\begin{eqnarray}
    P_\eq (x) = \frac{1}{Z} e^{- V(x)/k_B T} \, , \label{eq:thermal-equilibrium}
\end{eqnarray}
where $Z = \int_{-\infty}^{\infty} e^{- V(x)/k_B T} dx$ is the normalizing partition function. We can denote the excited contributions with an $*$ as
\begin{eqnarray}
    P^*(x,t) &=& e^{-V(x)/2 k_B T + V(0)/2 k_B T}  \sum_{k>0} N_k^{-2} \psi_k (x_0) \psi_k(x) e^{- \lambda_k t} \, . \label{eq:excited-states-PDF}
\end{eqnarray}
Initially, we place the system will be placed inside a box from $x=-L$ to $x=L$. This allows us to apply no-flux boundary conditions at $x=\pm L$.
As we will see, the eigenvalue spectrum for all the potentials $V(x)$ we will use throughout this Supplemental Material is continuous. Therefore, when we take the limit of $L\to \infty$, the sum can be replaced by an integral as
\begin{eqnarray}
    \sum_k = \int_0^\infty L \frac{dk}{\pi} \, .
\end{eqnarray}

\section{Long-time solution for $0 < \alpha < 1$}

As we will see, here we have anomalous relaxation of observables. We are interested in potentials that, for large $x$, obey $V(x) \approx V_0 |x|^\alpha$. To avoid the singularity at the origin that will emerge in the force $F(x) = - V'(x)$, we regularize the potential as
\begin{eqnarray}
    V(x) = V_0 \left(1 + \frac{x^2}{\ell^2} \right)^{\alpha/2} \, .
\end{eqnarray}
Henceforth we will scale the position variable by $\ell$  and correspondingly the time by $\ell^2/D$. For large values of $x$, the effective Schr\"odinger potential is
\begin{eqnarray}
    V_S(x) \approx \left( \frac{V_0}{k_B T} \right)^2 \frac{\alpha^2}{4 x^{2-2\alpha}} + \frac{V_0}{k_B T} \frac{\alpha(1-\alpha)}{2 x^{2-\alpha}} =  \frac{\beta^2 \alpha^2}{4 x^{2-2\alpha}} + \frac{\beta \alpha(1-\alpha)}{2 x^{2-\alpha}} \, ,
\end{eqnarray}
where, for simplicity, we defined $\beta \equiv V_0/k_B T$. The lowest eigenvalue, $\lambda_0 = 0$, will provide us with the Boltzmann-Gibbs thermal equilibrium state as in Eq.\,(\ref{eq:thermal-equilibrium}). Since the effective potential decays to zero at large distances, $V_S(x \to \pm \infty) \to 0$, the eigenfunctions in that region are equivalent to the free particle and the eigenvalues form a continuum. The excited states, $\lambda_k>0$, can be written as $\lambda_k = k^2$.

In the long-time limit, the largest contribution comes from the lowest eigenvalues. We will make use of the WKB solution,
\begin{eqnarray}
    \psi_k(x) = \left\{
    \begin{array}{lr} 
        A_k \frac{\exp \left( \int_x^{x_\TP} \sqrt{V_S(y) - k^2} dy \right)}{\left( V_S(x) - k^2 \right)^{1/4}} & \text{if } x < x_\TP \\
        2 A_k \frac{\sin \left( \int_x^{x_\TP} \sqrt{k^2 - V_S(y)} dy+ \pi/4\right)}{\left( k^2 - V_S(x)  \right)^{1/4}} & \text{if } x > x_\TP
    \end{array} \right. \, ,
\end{eqnarray}
where the turning point $x_\TP$ satisfies $V_S(x_\TP) = k^2$ and we will explain the role of $A_k$ shortly. We will focus our derivation in the $x>0$ region. Note that since we regularized the potential at the origin, there are in fact two turning points. We are only interested in the turning point furthest from the origin, which is not related to the regularization and can be found as the solution of
\begin{eqnarray}
    \frac{\beta^2 \alpha^2}{4 x_\TP^{2-2\alpha}} + \frac{\beta \alpha(1-\alpha)}{2 x_\TP^{2-a}} = k^2 \, .
\end{eqnarray}
The constant $A_k$ is obtained by the imposed boundary conditions at the origin. This constant will be different for the even eigenfunctions $\psi_k^\even(0) = 1$, ${\psi_k^\even} '(0) = 0$ and odd eigenfunctions $\psi_k^\odd(0) = 0$, ${\psi_k^\odd }'(0) = 1$. In the region near the origin and $x \ll x_\TP$ ($k^2 x^{2-\alpha} \ll 1$), we obtain perturbatively that
        \begin{eqnarray}
            \psi_k^{\even}(x) & \approx & e^{- \frac{V(x)}{2k_B T} + \frac{V(0)}{2k_B T} } \left( 1 - k^2 \int_0^x dy_1 \int_0^{y_1} e^{\frac{V(y_1)}{k_B T}- \frac{v(y_2)}{k_B T}} dy_2  \right) \approx - k^2 \frac{Z}{2} \frac{k_B T}{\alpha V_0 } x^{1-\alpha} e^{\frac{V_0 x^\alpha}{2k_B T} + \frac{V(0)}{2k_B T}} \label{eq:even-eigenfunction} \\
            \psi_k^{\odd}(x) & \approx & e^{- \frac{V(x)}{2k_B T} - \frac{V(0)}{2k_B T} } \int_0^x e^{\frac{V(y)}{k_B T}} dy \approx \frac{k_B T}{\alpha V_0}x^{1-\alpha} e^{\frac{V_0 x^\alpha}{2k_B T} - \frac{V(0)}{2k_B T}} \, , \label{eq:odd-eigenfunction}
        \end{eqnarray}
        where the integrals are replaced by their respective asymptotic limits.
        For small $k$ ($V_S(x) \gg k^2$), we may write the WKB prefactor as
    \begin{eqnarray}
        \left(  \frac{\beta^2 \alpha^2}{4 x^{2-2\alpha}} - k^2 \right) ^{-1/4} \approx \left( \frac{\beta^2 \alpha^2}{4 x^{2-2\alpha}} \right)^{-1/4} = \sqrt{ \frac{k_B T}{V_0} \frac{2 x^{1-\alpha}}{\alpha} } \, .
    \end{eqnarray}
    In our scaling of the exponent, we neglected non-leading terms that will be relevant to the pre-factor. We write (still in the $V_S(x) \gg k^2$ limit)
    \begin{eqnarray}
        \sqrt{V_S(x) - k^2} & \approx & \sqrt{\frac{\beta^2 \alpha^2}{4 x^{2-2\alpha}} + \frac{\beta \alpha(1-\alpha)}{2 x^{2-a}} - k^2} \nonumber \\
        &\approx & \sqrt{ \frac{\beta^2 \alpha^2}{4 x^{2-2\alpha}} - k^2 }  +  \frac{(1-\alpha)}{2x} \, ,
    \end{eqnarray}
    where the second line is obtained by considering that $x \gg 1$. The integral becomes
    \begin{eqnarray}
        \int_x^{x_\TP} \sqrt{V_S(y) - k^2}  dy \approx \sqrt{\frac{\pi}{4}} \left( \frac{\alpha V_0}{2k_B T} \right)^{\frac{1}{1-\alpha}} \frac{\Gamma\left( \frac{\alpha}{2 - 2 \alpha} \right)}{\Gamma\left( \frac{1}{2 - 2 \alpha} \right)} \, k^{-\frac{\alpha}{1-\alpha}} - \frac{\beta x^\alpha}{2} + \frac{(1-\alpha)}{2} \ln \left( \frac{x}{x_\TP} \right) \, .
    \end{eqnarray}
    We use that $x_\TP \approx (\alpha \beta / 2k)^{1/(1-\alpha)}$ to write that $(1-\alpha)\ln (x/x_\TP) \approx \ln (2 k x^{1-\alpha} / \alpha \beta) $. In this limit, the WKB solution has the same $x$ dependence as in Eqs.\,(\ref{eq:even-eigenfunction}) and (\ref{eq:odd-eigenfunction}), that is,
    \begin{eqnarray}
        \frac{\exp \left( \int_x^{x_\TP} \sqrt{V_S(y) - k^2} dy \right)}{\left( V_S(x) - k^2 \right)^{1/4}} \approx \frac{\sqrt{2 k}}{\alpha \beta} e^{-\sqrt{\frac{\pi}{4}} \left( \frac{\alpha V_0}{2k_B T} \right)^{\frac{1}{1-\alpha}} \frac{\Gamma\left( \frac{\alpha}{2 - 2 \alpha} \right)}{\Gamma\left( \frac{1}{2 - 2 \alpha} \right)} \, k^{-\frac{\alpha}{1-\alpha}}} x^{1-\alpha} e^{\frac{\beta x^\alpha}{2}} \label{eq:with-prefactor}
    \end{eqnarray}
        Matching the previous Eq. to  the solutions in Eqs.\,(\ref{eq:even-eigenfunction}) and (\ref{eq:odd-eigenfunction}), we obtain that
    \begin{eqnarray}
        A_k^{\odd} &=& \frac{1}{\sqrt{2k}} \exp \left\{ \sqrt{\frac{\pi}{4}} \left( \frac{\alpha V_0}{2k_B T} \right)^{\frac{1}{1-\alpha}} \frac{\Gamma\left( \frac{\alpha}{2 - 2 \alpha} \right)}{\Gamma\left( \frac{1}{2 - 2 \alpha} \right)} \, k^{-\frac{\alpha}{1-\alpha}} - \frac{V(0)}{2k_B T} \right\} \\
        A_k^{\even} &=& - \frac{Z \, k^{3/2}}{{2}^{3/2}} \exp \left\{ \sqrt{\frac{\pi}{4}} \left( \frac{\alpha V_0}{2k_B T} \right)^{\frac{1}{1-\alpha}} \frac{\Gamma\left( \frac{\alpha}{2 - 2 \alpha} \right)}{\Gamma\left( \frac{1}{2 - 2 \alpha} \right)} \, k^{-\frac{\alpha}{1-\alpha}} + \frac{V(0)}{2k_B T} \right\} \, .
    \end{eqnarray}
    Lastly, we must calculate the normalization $N_k$. Since we are interested in taking the $L \to \infty$ limit, the integral of $\psi_k(x)^2$ is going to be dominated by the large $x$ behavior, that is,
    \begin{eqnarray}
        N_k^2 = \int_{-L}^L dx  \,\psi_k(x)^2 \approx (2 A_k)^2 \int_{-L}^L dx \frac{\sin(k x + \pi/4)^2}{k} \approx \frac{(2 A_k)^2 L}{k} \, .
    \end{eqnarray}
    The summation can be replaced by an integral as
    \begin{eqnarray}
        \sum_{k>0} \frac{1}{N_k^2} = \int_0^\infty \frac{dk}{\pi} \frac{k}{4 A_k^2} \, .
    \end{eqnarray}

    \subsection{The exponent of the WKB solution at maximum $k^*$ is equivalent to the rate function}

    In the $L \to \infty$ limit, we can write Eq.\,(\ref{eq:excited-states-PDF}) as
    \begin{eqnarray}
        P^*(x,t) &=& \int_0^\infty \frac{dk}{\pi} \frac{k}{4 A_k^2} \psi_k(x) \psi_k(x_0) e^{-k^2t - \frac{V(x)}{2 k_B T} + \frac{V(0)}{2 k_B T}} \, .
    \end{eqnarray}
    First, we will consider the odd solution where $x_0$ is small enough so we may approximate ${\psi_k^\odd}(x_0) \approx x_0$ (from Eq.\,(\ref{eq:odd-eigenfunction})). In this case, we can write that
    \begin{eqnarray}
        P_\odd^* (x,t) &=& \int_0^\infty \frac{dk}{\pi} \frac{k}{4 A_k^2} \psi_k(x) \psi_k(x_0) e^{-k^2t - \frac{V(x)}{2 k_B T} + \frac{V(0)}{2 k_B T}}  \nonumber \\
        &=& x_0 \int_0^\infty \frac{dk}{\pi}  e^{-k^2t - \frac{V(x)}{2 k_B T} + \frac{V(0)}{2 k_B T}}
        \left\{
        \begin{array}{lr} 
        \frac{k}{4 A_k} \frac{\exp \left( \int_x^{x_\TP} \sqrt{V_S(y) - k^2} dy \right)}{\left( V_S(x) - k^2 \right)^{1/4}} & \text{if } x < x_\TP(k) \\
        \frac{k}{2 A_k} \frac{\sin \left( \int_x^{x_\TP} \sqrt{k^2 - V_S(y)} dy+ \pi/4\right)}{\left( k^2 - V_S(x)  \right)^{1/4}} & \text{if } x > x_\TP(k)
        \end{array} \right. \nonumber \\
        &=& x_0 e^{\frac{V(0)}{k_B T} }\int_0^\infty \frac{dk}{\pi}
        \left\{
        \begin{array}{lr} 
        \frac{\sqrt{2 k^3}}{4} \frac{\exp \left( - S_L(x,k) \right)}{\left( V_S(x) - k^2 \right)^{1/4}} & \text{if } x < x_\TP(k) \\
        \frac{\sqrt{2 k^3}}{2} \frac{\exp \left( - S_R(x,k) \right)}{\left( k^2 - V_S(x)  \right)^{1/4}} & \text{if } x > x_\TP(k)
        \end{array} \right. \label{eq:integral-PDF-odd}
        \, ,
    \end{eqnarray}
    where we grouped the exponentials contributions as
    \begin{eqnarray}
        S_L(x,k) &=& \sqrt{\frac{\pi}{4}} \left( \frac{\alpha V_0}{2k_B T} \right)^{\frac{1}{1-\alpha}} \frac{\Gamma\left( \frac{\alpha}{2 - 2 \alpha} \right)}{\Gamma\left( \frac{1}{2 - 2 \alpha} \right)} \, k^{-\frac{\alpha}{1-\alpha}}  + k^2 t + \frac{V(x)}{2k_B T} + \int_x^{x_\TP} \sqrt{V_S(y) - k^2} dy  \\
        S_R(x,k) &=& \sqrt{\frac{\pi}{4}} \left( \frac{\alpha V_0}{2k_B T} \right)^{\frac{1}{1-\alpha}} \frac{\Gamma\left( \frac{\alpha}{2 - 2 \alpha} \right)}{\Gamma\left( \frac{1}{2 - 2 \alpha} \right)} \, k^{-\frac{\alpha}{1-\alpha}}  + k^2 t + \frac{V(x)}{2k_B T} \pm i\int_{x_\TP}^x \sqrt{k^2 - V_S(y) } dy   \pm \frac{i \pi}{4}  \, .
    \end{eqnarray}
    As we are interested in the long-time limit, we can make use of the scaled variables
    \begin{eqnarray}
        z \equiv \frac{x}{t^{\gamma}} = \frac{x}{t^{\frac{1}{2-\alpha}}} \, \mathrm{ and } \, k \equiv \kappa t^{(\nu - 1)/2} = \kappa t^{(\alpha-1)/(2-\alpha)}
    \end{eqnarray}
    which allows us to write
    \begin{eqnarray}
        S_L(x,k) &\approx& t^\nu \left\{ \sqrt{\frac{\pi}{4}} \left( \frac{\alpha V_0}{2k_B T} \right)^{\frac{1}{1-\alpha}} \frac{\Gamma\left( \frac{\alpha}{2 - 2 \alpha} \right)}{\Gamma\left( \frac{1}{2 - 2 \alpha} \right)} \, \kappa^{-\frac{\alpha}{1-\alpha}}  + \kappa^2 + \frac{\beta z^\alpha}{2} + \int_z^{z_\TP} \sqrt{ \frac{\beta^2 \alpha^2}{4 y^{2-2\alpha}} - \kappa^2} \, dy \right\} \\
        S_R(x,k) &\approx& t^\nu \left\{ \sqrt{\frac{\pi}{4}} \left( \frac{\alpha V_0}{2k_B T} \right)^{\frac{1}{1-\alpha}} \frac{\Gamma\left( \frac{\alpha}{2 - 2 \alpha} \right)}{\Gamma\left( \frac{1}{2 - 2 \alpha} \right)} \, \kappa^{-\frac{\alpha}{1-\alpha}}  + \kappa^2 + \frac{\beta z^\alpha}{2} \pm i\int_{z_\TP}^z \sqrt{\kappa^2 - \frac{\beta^2 \alpha^2}{4 y^{2-2\alpha}}} \, dy  \right\}  \, .
    \end{eqnarray}
    For simplicity, we can define the exponent in these new variables as
    \begin{eqnarray}
        s_L(z,\kappa) & \equiv & \frac{S_L(x,k)}{t^\nu} \\
        s_R(z,\kappa) & \equiv & \frac{S_R(x,k)}{t^\nu} \, .
    \end{eqnarray}
    The integral in Eq.\,(\ref{eq:integral-PDF-odd}) can be calculated using Laplace's method (or stationary phase method). Specifically, the largest contribution to the integral arises from the vicinity of the point $k^*$ (or $\kappa^*$), which minimizes $s(z,\kappa)$. Up to pre-exponential factors, we have that
    \begin{eqnarray}
        P_\odd^* (x,t) \approx e^{-t^\nu s(z,\kappa^*)} \, ,
    \end{eqnarray}
    where $s(z,\kappa^*)$ can be either the left or right solution, depending on which part provides the largest contribution. It is clear then that this function, at the critical point, must be equivalent to the rate function
    \begin{eqnarray}
        \I(z) \equiv s(z,\kappa^*) \, .
    \end{eqnarray}
    We can show that this is true by proving that $s(z,\kappa^*)$ is a solution to the differential equation of $\I(z)$. Specifically, we must have that
    \begin{eqnarray}
        ( \partial_z s(z,\kappa^*) )^2 - \left( \frac{V_0}{k_B T} \frac{\alpha}{z^{1-\alpha}} + \frac{z}{2-\alpha} \right) \partial_z s(z,\kappa^*) + \frac{\alpha s(z,\kappa^*)}{2 - \alpha} \stackrel{?}{=} 0 \, . \label{eq:diff-eq-for-s}
    \end{eqnarray}
    Note that the minimum $\kappa^*$ is also a function of $z$. However, because it is a minimum, that is $\partial_\kappa s(z,\kappa)|_{\kappa = \kappa^*} = 0$, we can take the derivative with respect to $z$ as
    \begin{eqnarray}
        \partial_z [s(z,\kappa^*)] = \partial_z s(z,\kappa) \Big|_{\kappa = \kappa^*} + \partial_\kappa s(z,\kappa) \Big|_{\kappa = \kappa^*} = \partial_z s(z,\kappa) \Big|_{\kappa = \kappa^*} \, .
    \end{eqnarray}
    Using the previous equation and considering the left solution, after some manipulation, we write Eq.\,(\ref{eq:diff-eq-for-s}) as
    \begin{eqnarray}
      &=&  \sqrt{\frac{\pi}{4}} \left( \frac{\alpha V_0}{2k_B T} \right)^{\frac{1}{1-\alpha}} \frac{\Gamma\left( \frac{\alpha}{2 - 2 \alpha} \right)}{\Gamma\left( \frac{1}{2 - 2 \alpha} \right)} \, \alpha \kappa^{-\frac{\alpha}{1-\alpha}} - (1-\alpha) \kappa^2 \left\{ 1 + \int_{z}^{z_\TP} \frac{dy}{\frac{\beta^2 \alpha^2}{4 y^{2-2\alpha}} - \kappa^2}  \right\} \nonumber \\
      &=& (1-a)\kappa \left[ \sqrt{\frac{\pi}{4}} \left( \frac{\alpha V_0}{2k_B T} \right)^{\frac{1}{1-\alpha}} \frac{\Gamma\left( \frac{\alpha}{2 - 2 \alpha} \right)}{\Gamma\left( \frac{1}{2 - 2 \alpha} \right)} \, \frac{\alpha \kappa^{-\frac{1}{1-\alpha}} }{1 - \alpha} - \kappa \left\{ 1 + \int_{z}^{z_\TP} \frac{dy}{\frac{\beta^2 \alpha^2}{4 y^{2-2\alpha}} - \kappa^2}  \right\}  \right] \\
      &=& (1-a)\kappa \left[\partial_\kappa s(z,\kappa) \Big|_{\kappa = \kappa^*}  \right] = 0 \, .
    \end{eqnarray}
    In the last two lines of the previous equation, we identify the derivative with respect of $\kappa$. We see then that Eq.\,(\ref{eq:diff-eq-for-s}) is true at the maximum point and it is equivalent to the rate function.

    \subsection{Small $z$ solution for the PDF}

    For small $z$, the maximum $\kappa^*$ can be found on the left function $s_L$. 
    \begin{eqnarray}
        s_L(z,\kappa) &=& \sqrt{\frac{\pi}{4}} \left( \frac{\alpha V_0}{2k_B T} \right)^{\frac{1}{1-\alpha}} \frac{\Gamma\left( \frac{\alpha}{2 - 2 \alpha} \right)}{\Gamma\left( \frac{1}{2 - 2 \alpha} \right)} \, \kappa^{-\frac{\alpha}{1-\alpha}}  + \kappa^2 + \frac{z^\alpha}{2k_B T} + \int_z^{z_\TP} \sqrt{\frac{\beta^2 \alpha^2}{4 y^{2-2\alpha}} - \kappa^2} \, dy \nonumber \\
        & \approx &  \sqrt{\pi} \left( \frac{\alpha V_0}{2k_B T} \right)^{\frac{1}{1-\alpha}} \frac{\Gamma\left( \frac{\alpha}{2 - 2 \alpha} \right)}{\Gamma\left( \frac{1}{2 - 2 \alpha} \right)} \, \kappa^{-\frac{\alpha}{1-\alpha}}  + \kappa^2 + \kappa^2 \frac{k_B T}{V_0 \alpha(2-\alpha)} z^{2-\alpha} \, .
    \end{eqnarray}
    The $z$ dependent term in the previous equation is negligible for obtaining the maximum $\kappa^*$. The maximal point is then obtained 
    \begin{eqnarray}
        \partial_\kappa \left[ \sqrt{\pi} \left( \frac{\alpha V_0}{2k_B T} \right)^{\frac{1}{1-\alpha}} \frac{\Gamma\left( \frac{\alpha}{2 - 2 \alpha} \right)}{\Gamma\left( \frac{1}{2 - 2 \alpha} \right)} \,  \kappa^{-\frac{\alpha}{1-\alpha}}  + \kappa^2 \right]_{\kappa=\kappa^*} &=& 0 \nonumber \\
        - \sqrt{\pi} \left( \frac{\alpha V_0}{2k_B T} \right)^{\frac{1}{1-\alpha}} \frac{\Gamma\left( \frac{\alpha}{2 - 2 \alpha} \right)}{\Gamma\left( \frac{1}{2 - 2 \alpha} \right)} \, \frac{\alpha}{1-\alpha} {\kappa^*}^{-\frac{1}{1-\alpha}}  + 2{\kappa^*}  &=& 0 \, ,
    \end{eqnarray}
    leading to,
    \begin{eqnarray}
        \kappa^* &=& \left[ \sqrt{\frac{\pi}{4}} \left( \frac{\alpha V_0}{2k_B T} \right)^{\frac{1}{1-\alpha}} \frac{\Gamma\left( \frac{\alpha}{2 - 2 \alpha} \right)}{\Gamma\left( \frac{1}{2 - 2 \alpha} \right)} \, \frac{\alpha}{1-\alpha} \right]^{\frac{1-\alpha}{2-\alpha}} \, ,
    \end{eqnarray}
    and finally
    \begin{eqnarray}
        s_L(z,\kappa) &\approx& \frac{\left( \sqrt{\pi} \left( \frac{\alpha V_0}{2 k_B T} \right)^{\frac{1}{1-\alpha}} \frac{\Gamma\left( \frac{\alpha}{2 - 2 \alpha} \right)}{\Gamma\left( \frac{1}{2 - 2 \alpha} \right)} \right)^{1-\nu}}{\nu^\nu(1-\nu)^{1-\nu}} \left(1 + \frac{k_B T}{V_0 (2-\alpha)^2} z^{2-\alpha}  \right) = \Iodd(0) \left(1 + \frac{\nu k_B T}{V_0 \alpha(2-\alpha)} z^{2-\alpha}  \right) \, ,
    \end{eqnarray}
    where in the last line we identify the value of the rate function $\Iodd(0)$.
    The eigenfunction, in the $x \ll x_\TP$ limit, can be written in the scaled position variables as
    \begin{eqnarray}
        \psi_k(x) & \approx & \frac{x^{1-\alpha}}{\beta \alpha} e^{- \frac{V(0)}{2k_B T}} \exp \left\{ \frac{\beta x^\alpha}{2} - \frac{k^2 x^{2-\alpha}}{\beta \alpha (2-\alpha)} \right\} 
    \end{eqnarray}
    Combining all the information we obtained so far, we can find the PDF using Laplace's method to perform the integral to obtain
    \begin{eqnarray}
        P_\odd(x,t) & \approx & x_0 \, e^{ \frac{V(0)}{k_B T}}  \int_0^\infty \frac{d k}{2 \pi} \, k^2 \, \frac{x^{1-\alpha}}{\beta \alpha} e^{- \left\{ \sqrt{\pi} \left( \frac{\alpha V_0}{2k_B T} \right)^{\frac{1}{1-\alpha}} \frac{\Gamma\left( \frac{\alpha}{2 - 2 \alpha} \right)}{\Gamma\left( \frac{1}{2 - 2 \alpha} \right)} \, k^{-\frac{\alpha}{1-\alpha}} + k^2 t + \frac{k^2 x^{2-\alpha}}{\beta \alpha (2-\alpha)}\right\} } \nonumber \\
        & \approx & x_0 \left( \frac{k_B T}{\alpha V_0} x^{1-\alpha} \right) \frac{\nu \Iodd(0)}{t^{3/2-\nu}} \sqrt{\frac{\gamma - \nu}{4 \pi}} e^{\frac{V(0)}{k_B T}} \exp\left\{-\Iodd(0) t^\nu  - \frac{\Iodd(0)  k_B T x^{2-\alpha}}{V_0 (2-\alpha)^2 \, t^{1-\nu}} \right\} \, .
    \end{eqnarray}
    The pre-factor used in the main manuscript is immediately identified
    \begin{eqnarray}
       \mathcal{A} \left( t \right) &=& x_0 \left( \frac{k_B T}{\alpha V_0}  \right) \frac{\nu \Iodd(0)}{t^{3/2-\nu}} \sqrt{\frac{\gamma - \nu}{4 \pi}} e^{\frac{V(0)}{k_B T}} \, .
    \end{eqnarray}
    We may employ the same procedure to obtain the small-$z$ excited even PDF as
    \begin{eqnarray}
        P_\even(x,t) & \approx & - \int_0^{\infty} \frac{dk}{\pi} \frac{1}{Z} \frac{x^{1-\alpha}}{\beta \alpha} e^{- \left\{ \sqrt{\pi} \left( \frac{\alpha V_0}{2k_B T} \right)^{\frac{1}{1-\alpha}} \frac{\Gamma\left( \frac{\alpha}{2 - 2 \alpha} \right)}{\Gamma\left( \frac{1}{2 - 2 \alpha} \right)} \, k^{-\frac{\alpha}{1-\alpha}} + k^2 t + \frac{k^2 x^{2-\alpha}}{\beta \alpha (2-\alpha)}\right\} } \nonumber \\
        & \approx & - \frac{1}{Z} \left( \frac{k_B T}{\alpha V_0} x^{1-\alpha} \right) \sqrt{\frac{\gamma - \nu}{\pi t}}  \exp\left\{-\Ieven(0) t^\nu  - \frac{\Ieven(0)  k_B T x^{2-\alpha}}{V_0 (2-\alpha)^2 \, t^{1-\nu}} \right\} \, ,
    \end{eqnarray}
    and we identify the even pre-factor as
    \begin{eqnarray}
        \mathcal{B}(t) &=& - \frac{1}{Z} \left( \frac{k_B T}{\alpha V_0}  \right) \sqrt{\frac{\gamma - \nu}{\pi t}} \, .
    \end{eqnarray}
    The long-time prediction for the moments can be calculated using these expressions. For the odd moments we have
    \begin{eqnarray}
        \langle x^{2n-1} \rangle  & \approx & 2 \mathcal{A}(t) e^{-\Iodd(0) t^{\nu}} \int_0^\infty  x^{2n-\alpha}  e^{- \frac{\Iodd(0)  k_B T x^{2-\alpha}}{V_0 (2-\alpha)^2 \, t^{1-\nu}}} dx \nonumber \\
        & \approx &   \frac{2 \mathcal{A}(t) \Gamma \left( \frac{1+2n-\alpha}{2-\alpha} \right) e^{-\Iodd(0) t^{\nu}}}{(2-\alpha)\left( \frac{\Iodd(0)  k_B T}{V_0 (2-\alpha)^2 \, t^{1-\nu}} \right)^{\frac{1+2n-\alpha}{2-\alpha}}} \nonumber \\
        & \approx & \underbrace{ 2 x_0 \left( \frac{k_B T}{\alpha V_0}  \right) \nu \Iodd(0) \sqrt{\frac{\gamma - \nu}{4 \pi}} e^{\frac{V(0)}{k_B T}} \frac{\Gamma \left( \frac{1+2n-\alpha}{2-\alpha} \right) }{(2-\alpha)\left( \frac{\Iodd(0)  k_B T}{V_0 (2-\alpha)^2 } \right)^{\frac{1+2n-\alpha}{2-\alpha}}} }_{\mathcal{C}_n\left(\frac{V_0}{k_B T},\alpha \right)} \frac{e^{-\Iodd(0) t^{\nu}}}{t^{\frac{4(1-n)(1-\alpha)}{(2-\alpha)^2} + \frac{\nu^2}{2}} } \, ,
    \end{eqnarray}
    while for the even moments we have
    \begin{eqnarray}
        \langle x^{2n} \rangle  & \approx & 2 \mathcal{B}(t) e^{-\Ieven(0) t^{\nu}} \int_0^\infty  x^{2n+1-\alpha}  e^{- \frac{\Ieven(0)  k_B T x^{2-\alpha}}{V_0 (2-\alpha)^2 \, t^{1-\nu}}} dx \nonumber \\
        & \approx &   \frac{2 \mathcal{B}(t) \Gamma \left( \frac{2+2n-\alpha}{2-\alpha} \right) e^{-\Ieven(0) t^{\nu}}}{(2-\alpha)\left( \frac{\Ieven(0)  k_B T}{V_0 (2-\alpha)^2 \, t^{1-\nu}} \right)^{\frac{2+2n-\alpha}{2-\alpha}}} \nonumber \\
        & \approx & \underbrace{ - \frac{2}{Z} \left( \frac{k_B T}{\alpha V_0}  \right) \sqrt{\frac{\gamma - \nu}{\pi}} \frac{2 \mathcal{B}(t) \Gamma \left( \frac{2+2n-\alpha}{2-\alpha} \right) }{(2-\alpha)\left( \frac{\Ieven(0)  k_B T}{V_0 (2-\alpha)^2 } \right)^{\frac{2+2n-\alpha}{2-\alpha}}} }_{\mathcal{D}_n \left(\frac{V_0}{k_B T},\alpha \right)} \frac{e^{-\Iodd(0) t^{\nu}}}{t^{- \frac{4 + 3 \alpha^2 + 8n - 8\alpha (1+n)}{(2-\alpha)^2} } } \, .
    \end{eqnarray}